\documentclass[seceq]{ptptex}

\usepackage{graphicx}

\title{
Non-Abelian Strings in Hot or Dense QCD
}

\author{
Eiji \textsc{Nakano}$^a$\footnote{e.nakano(at)gsi.de}, 
Muneto \textsc{Nitta}$^b$\footnote{nitta(at)phys-h.keio.ac.jp, 
speaker at the conference}
and 
Taeko \textsc{Matsuura}$^c$\footnote{tamatsuu(at)east.ncc.go.jp}
}

\inst{
$^a$GSI, Planckstr.1, D-64291 Darmstadt, Germany \\
$^b$Department of Physics, Keio University, Hiyoshi, Yokohama,\\
Kanagawa 223-8521, Japan\\
$^c$Particle Therapy Division, Research Center for Innovative Oncology,\\ 
National Cancer Center Hospital East, Japan 
}

\abst{

Different types of non-Abelian vortex-strings 
appear in dense or hot QCD, both of which 
possess non-Abelian internal orientation zero modes. 
We calculate the interaction between them and find 
the universal repulsion for dense QCD (color superconductivity)
and the dependence on the orientations for hot QCD 
(chirally broken phase). 
This is a review article 
based on our papers 
\cite{Nakano:2007dq,Nakano:2007dr}.

}

\begin{document}

\maketitle

\section{Introduction and Summary}
It is one of the most important subjects to determine 
the phase structure of QCD 
in study of the strong interaction. 
QCD is expected to exhibit  
color superconductivity at high baryon density region,  
where the color $SU(3)_{\rm C}$, 
the flavor $SU(3)_{\rm F}$ 
and the baryon $U(1)_{\rm B}$
symmetries are spontaneously broken down to the 
the color-flavor locked (CFL) symmetry 
$SU(3)_{\rm C+F}$.\cite{Alford:2007xm}
According to the spontaneously broken 
$U(1)_{\rm B}$
there appear topological (vortex-)strings 
which wind around it.\cite{Forbes:2001gj,Iida:2002ev} 
These are global vortices \cite{Shellard:1987bv} or 
superfluid vortices in a rotating media. 
The existence of vortices in a core of a neutron star 
may become a proof of realization of 
the color superconductivity.
It has been shown\cite{Balachandran:2005ev} that 
there also exist non-Abelian strings 
which wind around both $U(1)_{\rm B}$ and 
$SU(3)_{\rm C-F}$ simultaneously 
because of the non-trivial homotopy 
$\pi_1 [U(3)] \simeq {\bf Z}$.\footnote{
Similar but different non-Abelian vortex-strings 
have been extensively studied in the context of 
supersymmetric gauge theories 
and superstring theory.\cite{Hanany:2003hp,Eto:2005yh}
Apart from supersymmetry, only but crucial difference 
is that $U(1)_{\rm B}$ is gauged in those context.  
}
The interesting is that these vortices 
carry a color gauge flux but behave as 
superfluid vortices in the energetic point of view.
Moreover 
they have normalizable orientational zero modes 
in the internal space, 
associated with further breaking of the residual symmetry 
$SU(3)_{\rm C+F}$ in the presence of the strings.\cite{Nakano:2007dr} 

On the other hand, topological strings 
are also expected to form during the chiral phase 
transition \cite{Zhang:1997is} 
where approximate axial symmetry $U(1)_{\rm A}$ 
is spontaneously broken 
as well as chiral symmetry $SU(3)_{\rm L} \times SU(3)_{\rm R}$ 
broken down to the vector-like symmetry $SU(3)_{\rm L+R}$.
These strings wind around the spontaneously broken $U(1)_{\rm A}$ 
which is explicitly broken by anomaly at low temperature 
but recovers at high temperature.  
Non-Abelian strings also exist due to  
$\pi_1[U(3)_{\rm A}] \simeq {\bf Z}$.
\cite{Balachandran:2002je,Nitta:2007dp}. 
These strings also have orientational zero modes 
in the internal space 
associated with breakdown of $SU(3)_{\rm L+R}$ 
in the presence of strings.
They are however non-normalizable\cite{Nakano:2007dq} unlike 
those appearing in the color superconductor.

Let us recall that, 
in the case of usual superconductors,  
their stability in the presence of external magnetic fields 
depends on whether 
topological (vortex-)strings attract or repel each other; 
the former is unstable (Type I) 
whereas the latter is stable (Type II). 
Therefore the purpose of our study is to calculate 
the interaction between two parallel non-Abelian strings 
with general relative 
orientational zero modes in the both cases of 
1) dense and 2) hot QCD. We find followings:\\
1)  In the case of dense
QCD, i.e. in the CFL phase, we find 
the universal repulsion between strings  
which does not depend on their orientations. 
This shows that 
the superfluid $U(1)_{\rm B}$ vortices \cite{Forbes:2001gj,Iida:2002ev}  
are unstable to decay, 
and that the most fundamental objects in the color superconductor is 
non-Abelian strings. 
Moreover it implies   
the stability of color superconductors 
in the presence of external color gauge fields.\\
2) 
In the case of hot QCD i.e. in the chirally broken phase,
we find that the force between strings  
depends on their relative orientations:
when the orientations are the same, 
the repulsive force reaches the maximum, 
whereas when the relative orientation becomes the maximum, 
no force exists.
This implies 
the marginal instability of  
$U(1)_{\rm A}$ strings\cite{Zhang:1997is}.
For more detail see the original papers 
\cite{Nakano:2007dq,Nakano:2007dr}.

\section{Non-Abelian Strings in Dense QCD}
We start from 
the most general 
Landau-Ginzburg(LG) Lagrangian 
\begin{eqnarray}\label{GL}
{\cal L} =
 {\rm{tr}}({D}\Phi)^{\dag}({D}\Phi)
-m^2{\rm{tr}} (\Phi^{\dag}\Phi)
-\lambda_1 ( {\rm{tr}}\Phi^{\dag}\Phi)^2
-\lambda_2 {\rm{tr}}\left[(\Phi^{\dag}\Phi)^2\right] 
-\frac{1}{4}F^a_{ij}F^{a   ij}, 
\end{eqnarray}
where $\Phi$ is a 3 by 3 matrix of scalar fields  
on which the symmetries act as
\begin{eqnarray}
\Phi \to e^{i\alpha} U_{\rm C} \Phi U_{\rm F}^{t}, \quad
 e^{i\alpha}  \in U(1)_{\rm B},\;\;  
 U_{\rm C} \in SU(3)_{\rm C}, \;\;
 U_{\rm F} \in SU(3)_{\rm F}.
\end{eqnarray}
In the vacuum $\langle \Phi \rangle = v {\bf 1}$ 
the symmetries are spontaneously broken down to the CFL symmetry 
$SU(3)_{\rm L+R}$. Hereafter we take $v=1$ for simplicity by 
a suitable rescaling.
In the polar coordinates $(\theta, \rho)$,  
a single non-Abelian string 
is given by \cite{Balachandran:2005ev}
\begin{eqnarray}\label{reference1}
&& \Phi(\theta, \rho)
 = {\rm diag}(e^{i\theta} f(\rho), g(\rho), g(\rho))
 \sim  {\rm diag}(e^{i\theta}, 1, 1) 
 \quad \mbox{for} \quad \rho \gg \lambda, \\
&& A_z = a(\rho)\, {\rm diag} (2,-1,-1) \sim 
 (1/ \rho)\, {\rm diag} (2,-1,-1)
\quad \mbox{for} \quad \rho \gg \lambda,
  \label{eq:sol-A}\
\end{eqnarray} 
where $f$ and $g$ are profile functions of $\rho$ 
behaving as $f\sim g \sim 1$ far from the core of a string  
($\rho \gg \lambda$) 
and $f\sim 0$ ($g\neq 0$) near the core of a string. 
Here the size $\lambda$ of the core is 
either the penetration depth or the coherent length.
A numerical solution with approximation $g=1$ 
can be found\cite{Balachandran:2005ev}. 
Note that the solution carries a color flux (\ref{eq:sol-A}) 
taking a value in the Lie algebra of $SU(3)_{\rm C}$. 

The solution (\ref{reference1}) breaks the CFL symmetry $SU(3)_{\rm C+F}$.
Therefore acting it  
on the solution (\ref{reference1}) 
we obtain a continuous family of solutions. 
Those solutions with different color flux one-to-one correspond to 
a complex projective space\cite{Nakano:2007dr}
\begin{eqnarray}
 {SU(3)_{\rm C+F} \over SU(2)_{\rm C+F} \times U(1)_{\rm C+F}} 
 \simeq {\bf C}P^{2} . \label{eq:orientation}
\end{eqnarray}

\begin{figure}
\begin{center}
\includegraphics[height=2.0cm, width=4cm]{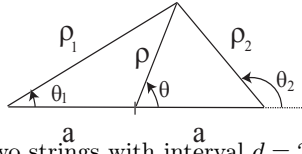}
\end{center}
\vspace{-5mm}
\caption{\label{fig1} 
Configuration of two strings with interval $d=2a$ 
in polar coordinate $(\rho, \theta)$.}
\end{figure}
Let us calculate the force between two strings 
with general orientations of ${\bf C}P^{2}$.  
First we put one string with a fixed orientation, 
given by the solution (\ref{reference1}) with $\theta\to\theta_1$ 
at one side of Fig.~\ref{fig1}. 
The other string put at the other side of Fig.~\ref{fig1}
should have generic orientation,  
which can be obtained in general by acting $SU(3)_{\rm F}$ 
on the form of the solution (\ref{reference1}) with $\theta\to\theta_2$. 
However in focusing on 
relative orientation between two strings,
we can consider an $SU(2)_{\rm F}$ subgroup 
(in the upper-left 2 by 2 block) without loss of generality:
$\left(\begin{array}{cc}
     e^{i\theta_2} f & 0 \\
                 0 & g \\  
   \end{array} \right)
 \left(\begin{array}{cc}
  a^* & -b \\
  b^* & a \\  
 \end{array} \right) 
= \left(\begin{array}{cc}
  e^{i\theta_2} a f & e^{i\theta_2} b f \\
             -b^* g & a^* g \\  
  \end{array} \right)$.
The point here is that this flavor rotated configuration reverts 
back to original one
at long distance by a color gauge rotation 
$\left(\begin{array}{cc}
a^* & -b e^{i \theta_2 F(\rho_2) }  \\
b^* e^{-i \theta_2 F(\rho_2)}  & a 
\end{array} \right)  \in SU(3)_{\rm C}$,
with an arbitrary regular function $F$ 
with conditions  
$F(0)=0$ and $F(\infty)=1$, 
to 
\begin{eqnarray}
\left(\begin{array}{cc}
  |a|^2 f e^{i \theta_2}+|b|^2 g e^{i\theta_2 F} 
& a^* b \left[-e^{i\theta_2 F}+f e^{i\theta_2} \right] \\
a b^* \left[-1+f e^{i(1-F)\theta_2} \right] 
&|a|^2 g+|b|^2 f e^{i(1-F)\theta_2 }   
 \end{array} \right)  
\sim 
\left(\begin{array}{cc}
 e^{i \theta_2} & 0 \\
 0 & 1 
 \end{array} \right) 
\; \mbox{for} \; \rho_2 \gg \lambda. \;\;\;\;\;\;
 \label{eq:transformed}
\end{eqnarray}
With this expression now the situation has become much simpler in
evaluation of long range force. 
The interaction energy of two strings, $E=\int F$, is
given by
subtracting energy of two isolated string systems from that 
of two-body system.
The Abrikosov's product ansatz with (\ref{eq:transformed}) results in
the energy density $F$ and the energy $E$: 
\begin{eqnarray}\label{F}
F = \pm
\frac{2}{3}
\left[ 
\frac{-a^2 + \rho^2}
{a^4 + \rho^4 -2 a^2 \rho^2 \cos (2 \theta)}
\right], \quad
E = \pm \frac{2 \pi}{3} 
\left[ -\ln 4 + \ln \{\left(a^2 + L^2 \right)/a^2\} \right],\;\;\;\;\;
\label{eq:int-energy}
\end{eqnarray}
with the infrared cutoff $L$.  
The upper(lower) sign denotes a 
string-(anti-)string configuration.
This gives the force 
between strings:\cite{Nakano:2007dr} 
\begin{eqnarray}\label{f}
 f(a,L)
 = {\mp} 
 \frac{\partial E}{2 \partial a}
 =  \pm \frac{2 \pi}{3} 
\left(\frac{1}{a} - \frac{a}{a^2 + L^2}  \right) 
\sim  \pm \frac{ 2 \pi }{3 a}, \label{eq:force1}
\end{eqnarray}
where the most-right side denotes 
the large volume limit $L \to \infty$.
Magnitude of the force looks shrunk by factor 1/3 from the
known result of Abelian global strings.\cite{Shellard:1987bv}

\section{Non-Abelian Strings in Hot QCD}
The LG Lagrangian (linear sigma model) for 
chiral symmetry breaking  is
\begin{eqnarray}\label{GL}
{\cal L}=
 {\rm{tr}} ({\vec \partial}\Phi^{\dag} {\vec \partial}\Phi)
-m^2{\rm{tr}} (\Phi^{\dag}\Phi)
-\lambda_1 ( {\rm{tr}}\Phi^{\dag}\Phi)^2
-\lambda_2 {\rm{tr}} [(\Phi^{\dag}\Phi)^2] 
\end{eqnarray}
where $\Phi$ is a 3 by 3 matrix of scalar fields  
on which the symmetries act as
\begin{eqnarray}
\Phi \to e^{i\alpha} U_{\rm L} \Phi U_{\rm R}^\dagger, \quad
 e^{i\alpha}  \in U(1)_{\rm A},\;\;  
 U_{\rm L} \in SU(3)_{\rm L}, \;\;
 U_{\rm R} \in SU(3)_{\rm R}.
\end{eqnarray}
These symmetries are spontaneously broken 
to the vector-like symmetry $SU(3)_{\rm L+R}$
by $\langle \Phi \rangle = v {\bf 1}$.
A single non-Abelian string 
with vanishing anomaly has the form
$\Phi(\theta, \rho)
 = {\rm diag}(e^{i\theta} f(\rho), g(\rho), g(\rho))$.
A numerical solution has been obtained.\cite{Nitta:2007dp}
This solution has orientation ${\bf C}P^2$ 
by replacing ${\rm C+F}$ in 
(\ref{eq:orientation}) 
by ${\rm L+R}$. 

We now calculate the inter-vortex force 
in the situation of Fig.~\ref{fig1}. 
The second string is of the form  
$g \Phi g^\dagger$ with 
$g = \cos \left({\alpha\over 2} \right) {\bf 1}_2 
   + i \vec n \cdot \vec{\sigma} \sin \left({\alpha \over 2}\right)$
with $\vec \sigma=(\sigma_1,\sigma_2,\sigma_3)$ the Pauli matrices 
and $\vec n$ a unit three vector.
Defining $e^{i \beta} = n_x - i n_y$, we have 
\begin{equation}\label{phi2}
g \Phi(\theta_2, \rho) g^\dagger =
\left(\begin{array}{cc}
e^{i\theta_2} \cos ^2 \left({\alpha\over 2}\right) 
 +  \sin ^2 \left({\alpha\over 2}\right)
& \frac{i}{2}(1-e^{i\theta_2}) e^{i \beta} \sin \alpha  \\
-\frac{i}{2}(1-e^{i\theta_2}) e^{- i \beta} \sin \alpha 
& \cos ^2 \left({\alpha\over 2}\right) 
 + e^{i\theta_2} \sin ^2 \left({\alpha\over 2}\right)  \\  
\end{array} \right).
\end{equation}
By using the Abrikosov's product ansatz again, 
we obtain the interaction energy density $\tilde{F}$ and 
integrated energy $\tilde{E}$:
$\tilde{F}=\pm 3(1+\cos \alpha)F/2$,
$\tilde{E}=\pm 3(1+\cos \alpha)E/2$, respectively.
Then the force between two strings is given by\cite{Nakano:2007dq}
\begin{eqnarray}
&& 
\tilde f(a,\alpha,L)= \mp  \frac{\partial \tilde E}{2 \partial a}
 = \pm (1 + \cos \alpha)  
\left(\frac{\pi}{a}
- \frac{\pi a}{a^2 + L^2}  \right)  
\sim 
\pm (1 + \cos \alpha) \frac{ \pi }{a}, 
\end{eqnarray}
where the most-right side denotes 
the large volume limit 
$L \to \infty$.  
Unlike the previous case (\ref{eq:force1}), this depends on
the relative orientation $\alpha$. 
For parallel orientations ($\alpha = 0$) this reaches 
the maximal coinciding with the Abelian case\cite{Shellard:1987bv}, 
whereas for orthogonal orientations ($\alpha = \pi$) this vanishes. 

\section*{Acknowledgements}
The work of E.N. is supported by 
Center for Theoretical Sciences, National Taiwan University 
under grant NO.~NSC96-2811-M-002-024. 
The work of M.N. is supported in part by Grant-in-Aid for Scientific
Research (No.~20740141) from the Ministry
of Education, Culture, Sports, Science and Technology.


%

\end{document}